# A Synthetic Error Analysis of Positioning Equation for Airborne Three-Dimensional Laser Imaging Sensor

Yuesong Jiang, Xiaoxia Liu, Ruiqiang Chen and Yanling Wang

School of Electronic and Information Engineering, Beijing University of Aeronautics and Astronautics, Beijing 100191, People's Republic of China

Email: yuesongjiang@vip.sina.com

**Abstract:**

This paper presents the exact error analysis of point positioning equation used for airborne three-dimensional (3D) imaging sensor. With differential calculus and principles of precision analysis a mathematics formula on the point position error and relative factors is derived to show how each error source affects both vertical and horizontal coordinates. A comprehensive analysis of the related error sources and their achievable accuracy are provided. At last, two example figures are shown to compare the position accuracy of elliptical trace scan and the line-trace scan under the same error source and some corresponding conclusions are drawn.

## 1 INTRODUCTION

Airborne LIDAR is composed of Global Positioning System (GPS), attitude measurement (i.e. inertial navigation system, INS) and laser scanning ranging, as utilized as a fast and efficient means of large scale DEM generation(Shan *et al* 2005), high quality ortho rectified image generation, quick response and assessment to flood risk, forest resource management(Koch *et al* 2006)and so on. Airborne LIDAR can also be a fast and accurate access to the ground three-dimensional data, which obtains geometric information of terrain and generates digital elevation model for quantification of three-dimensional remote sensing. As above, LIDAR systems are complex and multi-sensor systems, which result in five major kinds of error sources: individual sensor calibration, inter-sensor calibration, range measurement errors, time synchronization between system components and other errors including errors in the navigation solution (position and attitude errors). There have been many literatures on the error source description of airborne LIDAR, but most of them typically focus on a single or a few error sources and do not consider the combined effect of all error sources. Baltsavias (1999), for example, provides an overview of basic relations and error formulas concerning airborne laser scanning. Aloysius Wehr *et al* (1999) provide an introduction and overview of LIDAR. Vaughn *et al* (1996) develop the exact equation necessary to georeference the laser point, but he do not consider the error produced by the oscillating of the laser scan mirror. This paper provides a comprehensive analysis of the point positioning accuracy for the integrated LIDAR systems considering all the potential error sources by modifying the exact equation, with a view to the elliptical trace scan and the line-trace scan, consequently a reliable assessment of the achievable point positioning accuracy can be obtained. At last an example plotted of the achievable point positioning accuracies of different systems is shown and analyzed. The conclusions can be helpful to LIDAR manufacturers to avid or minimize some errors for LIDAR and to the LIDAR service providers to provide high quality LIDAR point clouds and products to users.

## 2  EXCACT POSITION EGUATION

*2.1 Referenced coordinate*

1) WGS-84 Cartesian reference frame $C$: The $X^C$-axis lies in the equatorial plane of the ellipsoid and passes through the ellipsoid's prime meridian; the $Z^C$-axis is the mean axis of rotation of the earth; the $Y^C$-axis completes a right-handed reference frame.

2) Body coordinate frame $B$: The body frame is rigidly attached to the aircraft with $X^B$-axis in the direction of flight, $Y^B$-axis is positive along the right wing of the flight, and $Z^B$-axis is in a downward direction perpendicular to the $X^B Y^B$ plane.

3) The laser beam coordinate frame $L$: The $Z^L$-axis points in the direction of the laser beam.

*2.2 The Scanning Trace*

In laser scanning, the distribution of the spots at which laser beam intersects the ground depends on the structure of the optical scanning system. Different structures lead to different scan path including the elliptical trace scan and the line trace scan, as shown in figure 1. In practical, the oscillating scanner and the rotating scanner can be used to achieve the line-scan. The elliptical scan can be completed by the palmer scanner.

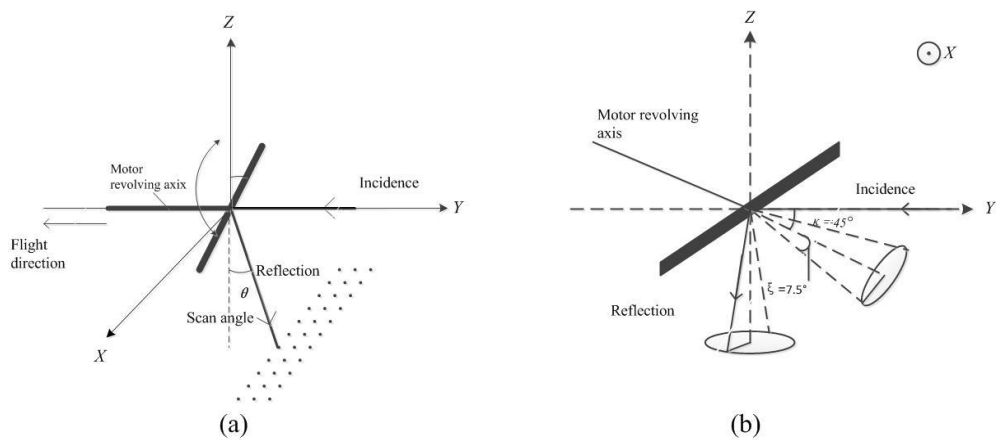

**Figure1.** (a) Schematic diagram of line-scan (b) Schematic diagram of elliptical-scan

*2.3 Exact positioning equation*

The calculation of ground coordinate of objects from scanning system observations has been well documented in the literature (Vaughn *et al* 1996). The coordinate on the ground can be calculated by combining the information from laser scanner, integrated GPS/INS navigation system and calibrated values. The basic geometry for the laser measurement is the vector relationships, as shown in figure 2.

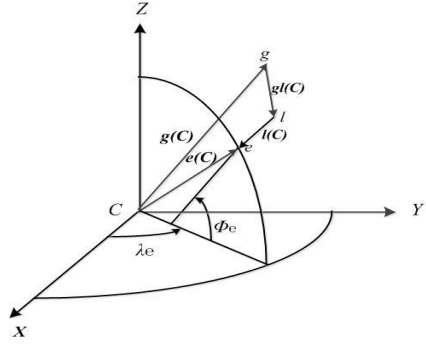

**Figure2.** Schematic diagram of integrated positioning system in WGS-84 reference frame

In short, the value of the GPS receiver $g$ which represents the kinematic position of the aircraft is provided by GPS data; from point $l$ the laser beam emits to obtain the range information from the laser scanner to the point $e$ where the laser beam intersects the earth. The three-dimensional coordinate of $e$ in WGS-84 $\mathbf{e(C)}$ is obtained using the data including the position and the attitudes (heading $\beta$, pitch $\alpha$, yaw $\varphi$ measured by INS) of the aircraft combined with the distance from $l$ to $e$ and the coordinate transformation, as following(Vaughn et al 1996):

$$\begin{bmatrix} X_e \\ Y_e \\ Z_e \end{bmatrix}^C = \begin{bmatrix} X_g \\ Y_g \\ Z_g \end{bmatrix}_C + \mathbf{R^{-1}}(\lambda, \phi - 90, 0)\mathbf{R^{-1}}(\beta, \alpha, \varphi) \cdot \left\{ \mathbf{R^{-1}}(\Delta\beta, \Delta\alpha, \Delta\varphi) \begin{bmatrix} 0 \\ 0 \\ |l(L)| \end{bmatrix}_L + \begin{bmatrix} X_{gl} \\ Y_{gl} \\ Z_{gl} \end{bmatrix}_B \right\} \quad (1)$$

Where $\mathbf{g(C)} = [X_g, Y_g, Z_g]_C^T$ is the coordinate of GPS receiving antenna in WGS-84; $\mathbf{l(L)} = [0, 0, |l(L)|]_L^T$ is defined in the laser beam coordinate, $|l(L)|$ obtained by the laser scanner, is the distance from laser firing point to ground point; $\mathbf{gl(B)} = [X_{gl}, Y_{gl}, Z_{gl}]_B^T$ is the offset vector from the GPS antenna to laser firing point in the body coordinate frame; $\lambda, \phi$ refer to the geodetic coordinates of latitude and longitude; if the laser points off-nadir when $\beta = \alpha = \varphi = 0$, we define the Euler angles $(\Delta\beta, \Delta\alpha, \Delta\varphi)$ of L rotated relative to B as counterclockwise rotations about $Z^L$.

In the inverse rotation matric $\mathbf{R}^{-1}(\kappa, \mu, \nu) = \mathbf{R}_3^{-1}(\nu) \cdot \mathbf{R}_2^{-1}(\mu) \cdot \mathbf{R}_1^{-1}(\kappa) = [\mathbf{R}_1(k) \cdot \mathbf{R}_2(\mu) \cdot \mathbf{R}_3(\nu)]^T$ $\kappa, \mu, \nu$ represent any matrix element. For an arbitrary angle of rotation $\sigma$, the following useful relationship applies:

$$\mathbf{R}_i^{-1}(\sigma) = \mathbf{R}_i^T(\sigma) = \mathbf{R}_i(-\sigma) \quad (2)$$

Where subscript $i$ refers to an arbitrary axis about which the rotation is made and the superscript $T$ designates the transpose of the matrix. The negative sign notes a clockwise direction rotation.

$$\mathbf{R}_1(\kappa) = \begin{bmatrix} 1 & 0 & 0 \\ 0 & \cos(\kappa) & \sin(\kappa) \\ 0 & -\sin(\kappa) & \cos(k) \end{bmatrix}$$

$$\mathbf{R}_2(\mu) = \begin{bmatrix} \cos(\mu) & 0 & -\sin(\mu) \\ 0 & 1 & 0 \\ \sin(\mu) & 0 & \cos(\mu) \end{bmatrix} \quad (3)$$

$$\mathbf{R}_3(\nu) = \begin{bmatrix} \cos(\nu) & \sin(\nu) & 0 \\ -\sin(\nu) & \cos(\nu) & 0 \\ 0 & 0 & 1 \end{bmatrix}$$

For the line trace scan, the scan mirror rotates back and forth in the plane which is perpendicular to the direction of the aircraft, but the changing velocity and acceleration of the mirror cause torsion between the mirrors, also the angular encoder and the point density increase at the edge of the scan field where the mirror slows down, and decreases at nadir.

Here, we transform the torsion into the rotation matrix $\mathbf{R}^{-1}(0 \quad 0 \quad \varphi_s)$, so the equation (1) can be modified as:

$$\begin{bmatrix} X_e \\ Y_e \\ Z_e \end{bmatrix}^C = \begin{bmatrix} X_g \\ Y_g \\ Z_g \end{bmatrix}_C + \mathbf{R}^{-1}(\lambda, \phi - 90, 0) \mathbf{R}^{-1}(\beta, \alpha, \varphi)$$

$$\cdot \left\{ \mathbf{R}^{-1}(0, 0, \varphi_s) \mathbf{R}^{-1}(\Delta\beta, \Delta\alpha, \Delta\varphi) \begin{bmatrix} 0 \\ 0 \\ |l(L)| \end{bmatrix}_L + \begin{bmatrix} X_{gl} \\ Y_{gl} \\ Z_{gl} \end{bmatrix}_B \right\} \quad (4)$$

Where $\varphi_s$ refers to the scan angle. After the computation based on Matrix Theory, we get

$$\begin{bmatrix} X_e \\ Y_e \\ Z_e \end{bmatrix} = \begin{bmatrix} X_g \\ Y_g \\ Z_g \end{bmatrix} + \begin{bmatrix} C_{11} & C_{12} & C_{13} \\ C_{21} & C_{22} & C_{23} \\ C_{31} & C_{32} & C_{33} \end{bmatrix} \cdot \left\{ \begin{bmatrix} A_{11} & A_{12} & A_{13} \\ A_{21} & A_{22} & A_{23} \\ A_{31} & A_{32} & A_{33} \end{bmatrix} \begin{bmatrix} 0 \\ 0 \\ |l(L)| \end{bmatrix} + \begin{bmatrix} X_{gl} \\ Y_{gl} \\ Z_{gl} \end{bmatrix} \right\} \quad (5)$$

where

$$C_{11} = \sin\phi\cos\alpha\cos\varphi - \sin\lambda\cos\phi\cos\alpha\sin\varphi + \cos\lambda\cos\phi\sin\alpha$$
$$C_{21} = \cos\lambda\cos\alpha\sin\varphi + \sin\lambda\sin\alpha$$
$$C_{23} = \cos\phi\cos\alpha\cos\varphi + \sin\lambda\sin\phi\cos\alpha\sin\varphi - \cos\lambda\sin\phi\sin\alpha$$
$$C_{21} = \sin\phi(\sin\beta\sin\alpha\cos\varphi - \cos\beta\sin\varphi) - \cos\lambda\cos\phi\sin\beta\cos\alpha$$
$$\quad - \sin\lambda\cos\phi(\sin\beta\sin\alpha\sin\varphi + \cos\beta\cos\varphi)$$
$$C_{22} = \cos\lambda(\sin\beta\sin\alpha\sin\varphi + \cos\beta\cos\varphi) - \sin\lambda\sin\beta\sin\alpha$$
$$C_{23} = \cos\phi(\sin\beta\sin\alpha\cos\varphi - \cos\beta\sin\varphi) + \cos\lambda\sin\phi\sin\beta\cos\alpha$$
$$\quad + \sin\lambda\sin\phi(\sin\beta\sin\alpha\sin\varphi + \cos\beta\cos\varphi)$$
$$C_{31} = \sin\phi(\cos\beta\sin\alpha\cos\varphi + \sin\beta\sin\varphi) - \cos\lambda\cos\phi\cos\alpha\cos\beta$$
$$\quad - \sin\lambda\cos\phi(\cos\beta\sin\alpha\sin\varphi - \sin\beta\cos\varphi)$$
$$C_{32} = \cos\lambda(\cos\beta\sin\alpha\sin\varphi - \sin\beta\cos\varphi) - \sin\lambda\cos\alpha\cos\beta$$
$$C_{33} = \cos\phi(\cos\beta\sin\alpha\cos\varphi + \sin\beta\sin\varphi) + \cos\lambda\sin\phi\cos\alpha\cos\beta$$
$$\quad + \sin\lambda\sin\phi(\cos\beta\sin\alpha\sin\varphi - \sin\beta\cos\varphi)$$

For the elliptical trace scan, the rotation matrix is $\mathbf{R}^{-1}(\beta_s, \alpha_s, \varphi_s)$, during which $\beta_s, \alpha_s, \varphi_s$ refer to the scan angles. The exact position equation of nutating scanner is expressed:

$$\begin{bmatrix} X_e \\ Y_e \\ Z_e \end{bmatrix} = \begin{bmatrix} X_g \\ Y_g \\ Z_g \end{bmatrix} + \begin{bmatrix} C_{11} & C_{12} & C_{13} \\ C_{21} & C_{22} & C_{23} \\ C_{31} & C_{32} & C_{33} \end{bmatrix} \cdot \left\{ \begin{bmatrix} B_{11} & B_{12} & B_{13} \\ B_{21} & B_{22} & B_{23} \\ B_{31} & B_{32} & B_{33} \end{bmatrix} \cdot \begin{bmatrix} 0 \\ 0 \\ |l(L)| \end{bmatrix} + \begin{bmatrix} X_{gl} \\ Y_{gl} \\ Z_{gl} \end{bmatrix} \right\}, \tag{6}$$

where

$$B_{13} = \cos(\varphi_s)\cos(\alpha_s)\left[\cos(\Delta\beta)\sin(\Delta\alpha)\cos(\Delta\varphi) + \sin(\Delta\beta)\sin(\Delta\varphi)\right]$$
$$\quad + \left[\cos(\beta_s)\sin(\alpha_s)\sin(\varphi_s) - \sin(\beta_s)\cos(\varphi_s)\right]$$
$$\quad \cdot \left[\sin(\Delta\beta)\sin(\Delta\alpha)\cos(\Delta\varphi) - \cos(\Delta\beta)\sin(\Delta\varphi)\right]$$
$$\quad + \left[\cos(\beta_s)\sin(\alpha_s)\cos(\varphi s) + \sin(\beta_s)\sin(\varphi_s)\right] \cdot \left[\cos(\Delta\alpha)\cos(\Delta\beta)\right]$$
$$B_{23} = \sin(\beta_s)\cos(\alpha_s)\left[\cos(\Delta\beta)\sin(\Delta\alpha)\cos(\Delta\varphi) + \sin(\Delta\beta)\sin(\Delta\varphi)\right]$$
$$\quad + \left[\sin(\beta_s)\sin(\alpha_s)\sin(\varphi_s) + \cos(\beta_s)\cos(\varphi_s)\right]$$
$$\quad \cdot \left[\sin(\Delta\beta)\sin(\Delta\alpha)\cos(\Delta\varphi) - \cos(\Delta\beta)\sin(\Delta\varphi)\right]$$
$$\quad + \left[\sin(\beta_s)\sin(\alpha_s)\cos(\varphi_s) - \cos(\beta_s)\sin(\varphi_s)\right] \cdot \cos(\Delta\alpha)\cos(\Delta\varphi)$$
$$B_{33} = -\sin(\alpha_s)\left[\cos(\Delta\beta)\sin(\Delta\alpha)\cos(\Delta\varphi) + \sin(\Delta\beta)\sin(\Delta\varphi)\right]$$
$$\quad + \cos(\alpha_s)\sin(\varphi_s)\left[\sin(\Delta\beta)\sin(\Delta\alpha)\cos(\Delta\varphi) - \cos(\Delta\beta)\sin(\nabla\varphi)\right]$$
$$\quad + \cos(\alpha_s)\cos(\varphi_s)\cos(\Delta\alpha)\cos(\Delta\varphi)$$

$$A_{13} = \cos(\varphi_s) \cdot \left[\cos(\Delta\beta)\sin(\Delta\alpha)\cos(\Delta\varphi) + \sin(\Delta\beta)\sin(\Delta\varphi)\right]$$
$$\quad - \sin(\varphi_s) \cdot \left[\cos(\Delta\beta)\sin(\Delta\alpha)\cos(\Delta\varphi) - \sin(\Delta\beta)\sin(\Delta\varphi)\right]$$
$$A_{23} = \sin(\varphi_s)\left[\cos(\Delta\beta)\sin(\Delta\alpha)\cos(\Delta\varphi) + \sin(\Delta\beta)\sin(\Delta\varphi)\right]$$
$$\quad + \cos(\varphi_s) \cdot \left[\cos(\Delta\beta)\sin(\Delta\alpha)\cos(\Delta\varphi) - \sin(\Delta\beta)\sin(\Delta\varphi)\right]$$
$$A_{33} = \cos(\Delta\alpha) \cdot \cos(\Delta\beta)$$

# 3 ERROR ANALYSIS

The three-dimensional coordinate of point $e$ is dependent on the following observed parameters shown in equation (4) and equation (6).

*3.1 Aircraft position offered by GPS/DGPS*

The kinematic location of the aircraft $(X_g, Y_g, Z_g)$ can be given by the simultaneously recording the GPS carrier phase and the code data from a ground base-station combined with the post-mission processing. In general, there are a number of factors, such as atmosphere errors, multipath, poor satellite geometry, and loss of the clock which have an influence on the accuracy of the positioning and are difficult to predict. Luckily, the kinematic positioning accuracy given by DGPS which is often used in literatures is that the position accuracy for short $(<30k\text{m})$ baseline is both $2c\text{m}$ horizontally and vertically, under the assumption of no loss of the clock of DGPS signals, the good satellite geometry and the minimal multipath (Glennie 2007). According to the statistics, all error sources considered, most airborne laser scanners' the position error ratio of the error in vertical to that in horizontal is 1.5 $(dZ_g/dX_g = dZ_g/dY_g = 1.5)$. In the simulation of this paper, we consider $dX_g = dY_g = 5c\text{m}, dZ_g = 7.5c\text{m}$ for 600 m flying height.

*3.2 Attitude measurement offered by INS*

The true heading, pitch, and roll angles of the aircraft $(\beta, \alpha, \varphi)$ are determined by a laser gyroscope (a commercial INS). The errors in LIDAR return position due to that the attitude errors are proportional to the range from the laser firing point to the intersecting point with the ground. In the process of simulation, these three Euler angles are taken as random angle value in a certain range. Based on the method of least square adjustment implemented by Terrapoint, the accuracies of 0.001 °in roll and pitch, and 0.004 °in heading have been achieved (Glennie 2007).

*3.3 Range measurement offered by laser scanner*

The real straight distance from the laser firing point to the intersecting point with the ground is defined as $l(L) = \left[0, 0, |l(L)|\right]^T$. Many factors affect the accuracy of $|l(L)|$, and can be divided into two categories: (1) the travel time spent during the round-trip between aircraft and the target point; (2) the uncertainty due to the laser beam divergence.

The travel time consists of two parts. One is that the index-of-refraction of the atmosphere leads to the slowing the speed of light and the atmospheric refraction causes the laser beam to curve along its path, according to Snell's law. It reaches a maximum at the edge of the scan and zero for a nadir scan. In order to minimize this kind of error source, the range measurement model should be corrected for considering the effects of the atmosphere. Our laser nominally points to the nadir, which allows us to ignore the correction due to Snell's law.

The other is the measurement of the travel time. An accurate range measurement made with a pulsed laser requires an accurate time interval measurement. A laser pulse starts, reflects from a surface, and is received by a detector. We start a timer at some consistent point on each transmitted pulse. The corresponding received pulse is recorded with a waveform digitizer that is triggered at some preset threshold by the leading edge of the pulse. All in all, the measurement of time is directly bound up with the ratio of the signal to noise, the threshold of the detector and the probability of false alarm.

The divergence of the laser beam gives rise to uncertainty in location of the actual point of the range measurement. For flat ground, the horizontal laser beam misalignment error causes the scanning line a smiley shape distortion across the track. The range measurement error can achieve the accuracy of $1\,c$m (May et al 2007).

*3.4 Lever-arm offset error*

The error consists of the vector from the GPS receiving antenna to the laser beam firing point of the scanner and the scan angle error produced by the oscillating of the scan mirror during the process of scanning. The vector can be got by a physical measurement in a majority of cases, under the assumption that the INS and the laser scanner are tightly aligned. The accuracy can achieve the accuracy of 0.2 centimeters in all three components. In order to make sure that the oscillating error is considered, a rotation matrix multiplies the $\mathbf{R^{-1}}(\Delta\beta, \Delta\alpha, \Delta\varphi)$ on the left side.

*3.5 Error produced by coordinate transformation*

The coordinate of GPS receiving antenna $\mathbf{g}(C)$ is recorded in the WGS-84, in order to get $\mathbf{R^{-1}}(\lambda, \phi-90, 0)$ we need to transform $\mathbf{g}(C)$ to the ellipsoidal coordinate. Generally speaking, the worst error in the latitude of $5 \times 10^{-7}$ degree occurs when the observer is three times the radium of the Earth.

*3.6 Error offered by the time synchronization*

The time synchronization between the various data streams of the integrated system is also a major error source. GPS receiver transmits one precise pulse per second, of which the leading edge resets a timer in the laser scanner, at the same time; it also affects the time lags of INS. The time synchronization leads to the low-grade of the variables: (1) the range $l(L)$'s measurement; (2) the measurement of the attitude angles. The clock offsets and drifts must be determined through the calibration with an accurate integration. Any time delays must be known, or they will introduce errors in the integration.

## 4 ACCURACY ESTIMATE

In practice, flat terrain is assumed; sloped terrain will cause additional errors in the vertical coordinate. For the line trace scan, by substituting $A1, A2, A3$ into equation (5), with differential calculus and principles of precision analysis a set of mathematics formula on the

related error sources is derived as following:

$$\begin{aligned}
dX_e &= dX_g + X_{gl}^B dC_{11} + M_{12} dC_{12} + M_{13} dC_{13} \\
&\quad + M_{14} d\varphi_s + M_{15} dl + C_{11} dX_{gl}^B + C_{12} dY_{gl}^B + C_{13} dZ_{gl}^B \\
dY_e &= dY_g + X_{gl}^B dC_{21} + M_{12} dC_{22} + M_{13} dC_{23} \\
&\quad + M_{24} d\varphi_s + M_{25} dl + C_{21} dX_{gl}^B + C_{22} dY_{gl}^B + C_{23} dZ_{gl}^B , \\
dZ_e &= dZ_g + X_{gl}^B dC_{31} + M_{12} dC_{32} + M_{13} dC_{33} \\
&\quad + M_{34} d\varphi_s + M_{35} dl + C_{31} dX_{gl}^B + C_{32} dY_{gl}^B + C_{33} dZ_{gl}^B
\end{aligned} \quad (7)$$

where $dX_e, dY_e, dZ_e$ are the errors in the three coordinate, among which

$$\begin{cases}
M_{12} = Y_{gl}^B - \sin(\varphi_s) \cdot l \\
M_{13} = Z_{gl}^B + \cos(\varphi_s) \cdot l \\
M_{i4} = -[C_{i3} \cdot \sin(\varphi_s) - C_{i2} \cdot \cos(\varphi_s)] \cdot l , \\
M_{i5} = C_{i3} \cdot \cos(\varphi_s) - C_{i2} \cdot \sin(\varphi_s) \\
dC_{ij} = \dfrac{\partial C_{ij}}{\partial \beta} d\beta + \dfrac{\partial C_{ij}}{\partial \alpha} d\alpha + \dfrac{\partial C_{ij}}{\partial \varphi} d\varphi
\end{cases} \quad (8)$$

The mean-square error is

$$dR = \left(dX_e^2 + dY_e^2 + dZ_e^2\right)^{\frac{1}{2}} . \quad (9)$$

From the equation (7), obviously, the point position accuracy at which laser beam interests the ground depends on the every error source discussed above. Under the assumption that all the error sources are relatively independent, variance propagation law is used to calculate the total errors of the three coordinate, then mean-square error of $dX_e, dY_e, dZ_e$ is derived as the accuracy assessment value.

For the elliptical trace scan, the equation (6) can be simplified and expressed as

$$\begin{bmatrix} X_e \\ Y_e \\ Z_e \end{bmatrix} = \begin{bmatrix} X_g \\ Y_g \\ Z_g \end{bmatrix} + \begin{bmatrix} C_{11} & C_{12} & C_{13} \\ C_{21} & C_{22} & C_{23} \\ C_{31} & C_{32} & C_{33} \end{bmatrix} \cdot \left\{ |l(L)| \cdot \begin{bmatrix} -\sin(2\delta)\sin(\gamma) \\ \cos(2\delta) \\ \sin(2\delta)\cos(\gamma) \end{bmatrix} + \begin{bmatrix} X_{gl} \\ Y_{gl} \\ Z_{gl} \end{bmatrix} \right\}, \quad (10)$$

where

$$\gamma = tg^{-1}\left[\frac{\sin\xi \cdot \sin\theta}{\cos\kappa \cdot \sin\xi \cdot \cos\theta - \sin\kappa \cdot \cos\xi}\right],$$

$$\delta = \cos^{-1}\left(\sin\kappa \cdot \sin\xi \cdot \cos\theta + \cos\kappa \cdot \cos\xi\right).$$

The angle between $Y-axis$ and the motor revolving axis $\xi$ is 45degree and the angle between the palmer scan mirror and the motor revolving axis $\kappa$ is 7.5degree, as shown in figure 1 (b). The motor revolving angle is $\theta$. To illustrate the effect of the accuracy of all the error parameters, the same options as preceded are taken. The error distribution of the line trace scan on the left side and elliptical trace scan on the right side are shown in figure 3.

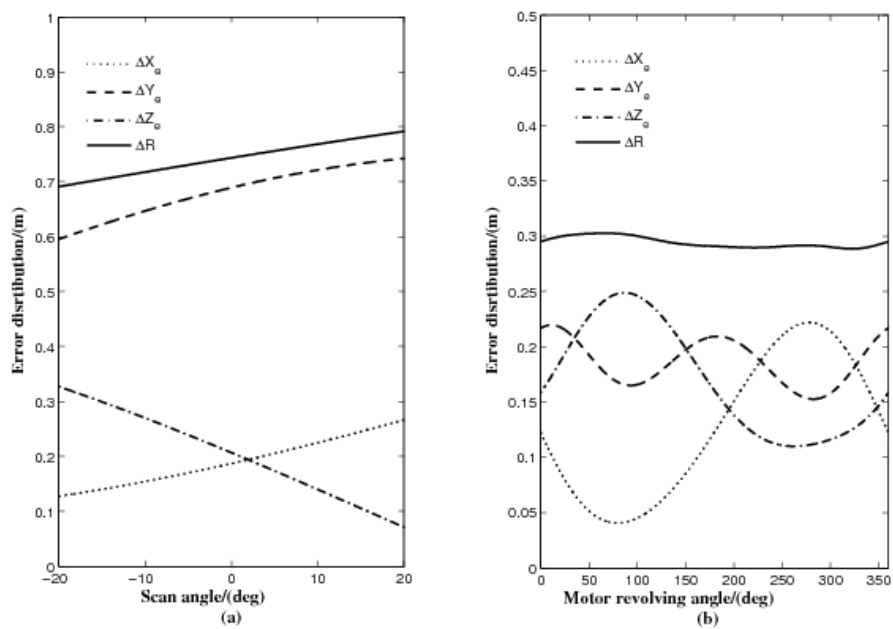

**Figure3.** Effect of all errors on point positioning for L=600m

Apparently, the elliptical trace scan causes relatively smaller error compared with the line trace scanner. For the line trace scanner, even though the scan angle rotates back and forth symmetrically, none of the errors is symmetrical; The error of $X-axis$ and $Y-axis$ have the same upward tendency, otherwise, the error of $Z-axis$ decreases; The mean-square error is mostly up to the error of $Y-axis$. For the elliptical scanner, the error of three-dimensional coordinate has waves whose waveform resemble a sine curve; The mean-square error is stable at the accuracy of 0.3m with slight fluctuations.

## 5 CONCLUSIONS

This paper presents a comprehensive error source analysis based on the exact point position accuracy equation. The airborne integrated system is really complicated. With differential calculus and principles of precision analysis, a set of rather realistic and reliable accuracy polynomial related to all the major error sources is derived to show how each error affects the positioning accuracy. To achieve the accuracy of the point position, the improvement of GPS position, the attitude and the range measurement must be made. Lever-arm offset and the integrated error need to be minimized. In addition, the LIDAR system should be well post-processing in order to eliminate the clock error.